# Unveiling the influence of behavioural, built environment and socio-economic features on the spatial and temporal variability of bus use using explainable machine learning

Sui Tao[a], Francisco Rowe[b*], Hongyu Shan[c]

[a]*Faculty of Geographical Science, Beijing Normal University, Beijing, China*
[b]*Department of Geography and Planning, The University of Liverpool, Liverpool, UK*
[c]*Faculty of Geographical Science, Beijing Normal University, Beijing, China*

* Corresponding author: F.Rowe-Gonzalez@liverpool.ac.uk

## Acknowledgements

This work was supported by the National Natural Science Foundation of China (No. 42201217).

**Unveiling the influence of behavioural, built environment and socio-economic features on the spatial and temporal variability of bus use using explainable machine learning**

**Abstract**

Understanding the variability of people's travel patterns is key to transport planning and policy-making. However, to what extent daily transit use displays geographic and temporal variabilities, and what are the contributing factors have not been fully addressed. Drawing on smart card data in Beijing, China, this study seeks to address these deficits by adopting new indices to capture the spatial and temporal variability of bus use during peak hours and investigate their associations with relevant contextual features. Using explainable machine learning, our findings reveal non-linear interaction between spatial and temporal variability and trip frequency. Furthermore, greater distance to the urban centres (>10 kilometres) is associated with increased spatial variability of bus use, while greater separation of trip origins and destinations from the subcentres reduces both spatial and temporal variability. Higher availability of bus routes is linked to higher spatial variability but lower temporal variability. Meanwhile, both lower and higher road density is associated with higher spatial variability of bus use especially in morning times. These findings indicate that different built environment features moderate the flexibility of travel time and locations. Implications are derived to inform more responsive and reliable operation and planning of transit systems.

**Keywords**: Public Transit, Travel Behaviour, Behavioural Variability, Smart Card Data, Machine Learning

## 1. Introduction

Public transport serves as a key component of a city's transport system, the operation of which is important for the various social and economic functions of our societies (Vuchic, 2005). However, public transport tends to be less flexible and less comfort compared to private vehicles. As the level of car ownership worldwide grows, poor transit service coupled with unfriendly environment (e.g., low density, homogeneous land use) may quickly translate into loss of ridership (Merlin et al., 2021; Tao et al., 2019). Facing such challenges, it is critical for transit agency and service to better understand and accommodate the various travel needs of transport passengers. In this regard, capturing the spatial-temporal dynamics of people's transit use and understanding the meridian of factors contribution to influence travel behaviour remain a primary and necessary step towards establishing more responsive and attractive transit system (Morency et al., 2007; Tao et al., 2014).

Understanding the variability of people's travel patterns across transport networks provides essential input for transport modelling, marketing and policy-making (Deschaintres et al., 2022; Susilo and Axhausen, 2014). Instead of being entirely fixed in space and time, people's daily mobility is versatile in nature often involving irregular shifts in travel time, locations and routes (Buliung et al., 2008; Egu and Bonnel, 2020). Hence, developing a full grasp of the spatial and temporal variability of travel behaviour has emerged as a critical dimension for transport modelling, transit operation and customer analysis (Goulet-Langlois et al., 2016; Kim et al., 2017). Traditionally, scholars have used household travel survey to capture and examine interpersonal and intrapersonal variability of daily mobility (e.g., Hanson and Huff, 1986; Buliung et al., 2008; Schlich and Axhausen, 2003; Susilo and Axhausen, 2014). Their findings indicated the presence of fluctuation of travel behaviour at the individual level (e.g., variations of trip time and mode choice), reflecting constantly varying travel demand. Moreover, scholars also affirmed significant variability particularly for people's mandatory trips (e.g., varying commuting times and locations) on weekdays, partially attributed to both the surrounding environment (e.g., land use patterns, provision of transit service) and different social and household roles of individuals (e.g., Crawford, 2020; Shen et al., 2013).

Over recent years, the emergence of smart card data (SCD) has prompted new waves of research on travel behaviour dynamics. Compared to survey data, SCD provide richer information of people's transit trips often offering large volume of observations and fine spatial-temporal scales (Munizaga and Palma, 2012; Pelletier et al., 2011; Tao et al., 2021). Given such features of SCD, scholars have probed in greater depth the variability of people's transit use (e.g., whether or not travelling by fixed routes and time), hoping to better unveil whether there exist any discernible patterns and driving factors that may explain the irregularity of daily mobility (Egu and Bonnel, 2020; Kim et al., 2017). To do so, sophisticated data-mining approaches have been adopted to extract the different types of transit use patterns, which, in turn, has been used to identify transit passengers associated with distinctive spatial-temporal travel dynamics (Goulet-Langlois et al., 2016; Kieu et al., 2015). Various measures (such as spatial-temporal similarity index) and new concepts (such as stickiness) have also been developed to capture and evaluate the spatial and temporal regularity and irregularity of transit use over time (Ma et al., 2018; Wei, 2022). The extracted information can be used to inform the operation and planning of transit service (e.g., adjusting transit routes, rescheduling, providing customised information) to better anticipate and cater for changing transit demands.

However, a scrutiny of the literature indicates that the variability of individual transit use has not been fully understood. In particular, while some previous research has shed light on the spatial and temporal dynamics in isolation, limited research has sought to understand the variability of these dimensions and disentangled their influencing factors simultaneously (Egu and Bonnel, 2020). Some fundamental questions remain unclear:

(1) How may the spatial and temporal variability of transit vary across urban space and time periods (e.g., morning *versus* evening peak hours)?
(2) To what extent do different social-spatial and behavioural features act to shape the spatial and temporal variability in transit use? and,
(3) What is the shape of relationships governing the influence of behavioural, built environment and socio-economic features on the spatial and temporal variability of transit use?

Addressing these questions will help developing our understanding of systematic differences in people's diversified demands for transit service and their underlying mechanisms.

To address the research questions above, we will adopt a suite of simple yet novel measures to capture the spatial and temporal variability of transit use, and examine its spatial patterns during peak hours. We will use XGBoost (eXtreme Gradient Boosting) to quantify the relationship between transit use variability and behavioural, built environment and socio-demographic features. We will also use SHapley Additive exPlanations (SHAP) values to identify and understand the non-linear relationships between the transit use variability and relevant features.

The remainder of the study is structured as follows. Section 2 provides a brief review of the relevant literature, before identifying the knowledge gaps. Section 3 introduces the study context, data sources and methods employed. Section 4 reports the analytical results. Section 5 discusses the main findings and policy implications, before conclusions are drawn in Section 6.

## 2. Literature review

### 2.1 The variability of daily travel behaviour

Understanding the dynamics of daily travel behaviour (e.g., trip frequency, trip distance and duration, departure time and destination choice) has been fundamental to transport research and planning (Schönfelder and Axhausen, 2016). Conventionally, research in this area has largely sought to elucidate interpersonal and intrapersonal variability of travel behaviour based on household travel survey. A well-known example is the 5-weeky travel diary in Uppsala, Sweden. Using this dataset, Hanson and Huff (1986) identified five groups with different travel behavioural patterns and socio-demographic attributes, although notable intra-group variation was also detected. In addition, they revealed that the fluctuation of travel behaviour (e.g., activity, travel mode, travel time and location) was common on both daily and weekly bases (Huff and Hanson, 1986; Hanson and Huff, 1988). Similar findings have also been reported in other contexts including the UK (e.g., Pas and Koppleman, 1987) and the US (e.g., Lockwood et al., 2005).

Following earlier works, scholars have further investigated intrapersonal travel behaviour with increasing interest on discerning activity types and patterns. For example, drawing on a six-week travel diary in Germany, Schlich and Axhausen (2003) confirmed that people's daily travel behaviour was neither '*totally repetitious nor totally variable*'. Schlich et al. (2004) further unveiled that travel behaviour was plausibly more variable on weekends than on weekdays. Focusing on Toronto, Buliung et al. (2008) employed a series of measures to describe the spatial properties of people's activity-travel behaviour (e.g., geographical extent, activity dispersion). Their findings confirmed marked weekday-to-weekend transition of activity-travel behaviour. In another related study, Susilo and Axhausen (2014) employed Herfindahl-Hirschman Index (a market concentration metric) to examine people's choices of daily activity-travel-location combinations, the variation of which was found to be conditioned on out-of-home commitment, household conditions and accessibility of activity locations, but less linked to modal choice. Furthermore, using a 5-week GPS-based survey data, Watanabe et al. (2021) compared the time use dynamics of two small samples (n=21 and 50) in Japan. Apart from intrapersonal variation, they found that the individual time-use patterns varied across different cities, which was partially attributed to their varied transport infrastructures and land use patterns.

Furthermore, some scholars paid particular attention to probing the variability of commuting, which has been conventionally assumed to be rather fixed over space and time. For example, drawing a 7-day GPS dataset, Shen et al. (2013) analysed the commuting patterns of a small sample (n=96) and revealed that individuals' commuting trips were highly flexible, constantly changing in terms of travel time, location, mode and route. In addition, Crawford (2020) extracted spatial and temporal features of commuting using the UK National Travel Survey data over 19 years. Through clustering and regression analysis, Crawford identified four groups (e.g., infrequent, spatially variable, temporally variable and regular) with distinctive commuting dynamics, the distribution of which were found to vary over socio-demographic groups (e.g., gender, age) and time. Their findings highlighted the need of considering the increasingly diversified commuting needs.

The aforementioned studies provided valuable insights into the variability of daily travel behaviour. However, compared to conventional indicators (e.g., trip frequency and duration), variability has still not received as much attention, despite its implications for better predicting varying travel demand and planning more responsive transit services. In addition, most existing research drew on multiday travel survey. Issues such as sample bias, inaccurate information, coarse spatial and temporal scales were often present that prevented more in-depth investigation of different modal use (Deschaintres et al., 2022). Moreover, it appears challenging to extend the observed variability of activity-travel behaviour to the public transport context, given that transit use information in the traditional travel survey can be relatively limited. Considering that public transport constitutes a key component of today's urban transport systems, capturing and understanding people's dynamics of transit use warrants further attention.

### 2.2 Smart card data and transit use variability

Over recent years, smart card data (SCD) have gained increasing interest. SCD usually store large amount of real-time transit trip records that were normally unavailable in traditional household travel survey (Pelletier et al., 2011; Tao et al., 2021). The emergence of SCD has prompted investigations of the spatial and temporal variability of travel behaviour within the public transport context. In this area, a group of research employed data-mining coupled with geo-visual techniques to extract and examine different behavioural dimensions, including boarding and alighting behaviour (Song et al., 2018), transfer behaviour (Ma et al., 2018), activity anchors (Chu and Chapleau, 2010), passenger flow and travel trajectories (Tao et al., 2014; Cheng et al., 2021) and the impacts of weather on bus ridership (Tao et al., 2018).

Given the advantages associated with smart card data, substantial research has examined the dynamics of transit use via sophisticated methods and algorithms. A main strand of research employed machine-learning approaches to classify transit passengers with regular and irregular spatial-temporal travel features (He et al., 2021; Ma et al., 2013). For example, drawing on three months of SCD in London, Manley et al. (2018) employed density-based spatial clustering of applications with noise (DBSCAN) to extract high-density (or repeated) travel events for individual bus and rail passengers, which was used to represent the regularity of travel patterns. Through visualisation exercises, they revealed the distribution of regular journeys over different times of day and across the spatial context. Similar analysis of regularity detection and clustering has also been carried out in Brisbane, Australia (Kieu et al., 2015) and Japan (Liu et al., 2022). In another related study, using SCD in Nanjing, Lei et al. (2020) developed an algorithm to identify 'temporal motifs' (or typical travel episodes), which was applied to categorise individual travel patterns of metro and bikesharing users.

Some scholars have sought to more explicitly measure the level of variability of individual-level transit use. For example, using five days of SCD, Ma et al. (2013) extracted four regularity features (i.e., travel days, number of similar first boarding times, number of similar route sequences and number of similar stop ID sequences). Also by applying clustering analysis, they identified passenger groups with different levels of transit use regularity. Using one-month SCD in London, Goulet-Langlois et al. (2016) constructed an entropy rate indicator to capture and categorise transit riders' activity and travel sequence, which was shown to be useful in capturing irregular activity patterns that did not align with calendar periodicity. However, spatial details (e.g., travel routes, locations) were not captured in the proposed measure. Using a 6-month SCD in Brisbane, Australia, Kim et al. (2017) developed a stickiness index to quantify to what extent bus riders might or not stick to fixed routes among all available route options between given OD pairs. Further regression analysis indicated that user characteristics (e.g., trip frequency), OD and journey features (e.g., travel time, direction and time period) were significantly related to the stickiness of route choice. Also focusing on Brisbane, Wei (2022) constructed a trip-level similarity index that incorporated trip origin, destination and timestamps, which was found to be significantly affected by short-term weather conditions (e.g., rainfall, wind speed).

Furthermore, scholars have combined interpersonal and intrapersonal perspective to obtain more compressive insights of transit use variability. Based in Lyon, France, Egu and Bonnel (2020) refined a traditional metric to represent the spatial and temporal similarity of individual transit use. Through clustering analysis, they further identified groups with discernible levels of variability in terms of the regularity, trip duration and intensity of transit use. However, partially due to data limitations, their similarity measurement was relatively coarse, using time periods spanning hours and districts only for boarding stops. Lastly, using one-year SCD in Shizuoka, Japan, Liu et al. (2021)

analysed the role of age in travel pattern variability captured by clustering analysis and entropy measures. Their findings highlighted that younger old group (aged between 65-74) exhibited lower day-to-day variability of transit use than the older ones.

Despite an accumulating body of research, the current understanding of transit use dynamics remains incomplete. First, limited research has simultaneously examined the spatial and temporal variability of transit use (i.e., whether or not travelling by fixed locations and times). Among those tapping on both dimensions, certain details were not captured (e.g., alighting stop and time) (e.g., Ma et al., 2013; Egu and Bonnel, 2020). Hence, it remains unclear that to what extent the two dimensions tend to co-occur. Second, it remains unclear how spatial and temporal variability of transit use may be influenced by different external factors. As noted above, it has been evidenced that various behavioural (e.g., trip frequency, duration) and built environment attributes (e.g., land use patterns) can contribute to transit use variability (e.g., using different routes or stops) through influencing people's travel habit and ease of accessing different transport and activity opportunities (Kim et al., 2017; Liu et al., 2022). However, it is less known whether they may exert differentiated effects on the spatial and temporal variability of transit use. Addressing these issues will arguably enhance our understanding of the versatile nature of travel behaviour and help make more targeted interventions accordingly.

## 3. Methodology

### 3.1 Study context

Beijing, the capital city of China, is the study context. The population of Beijing is over 20 million and its land area over 16,410 $km^2$ (Figure 1). The city's public transport has witnessed substantial expansion over the past couple of decades. At present, it has 23 metro lines and over 1500 bus routes in operation. For the current study, the area within the $6^{th}$ Ring Road is the focus, which consists of the core urban districts and the inner suburban areas of the city. We also focus on bus trips which account for over 50% of all transit trips. Furthermore, the study area was divided into 500 by 500 metre grids, which is used as the basis to spatially join the individual trip information and socio-demographic data (as described below). Previous research showed that individuals tend to walk relatively short distances to reach bus stops, typically within the range of 300 to 500 metres (e.g., El-Geneidy et al., 2014; Tennøy et al., 2022). Hence, we consider the adoption of a 500 metre grid is justifiable. The centroids of the grids were employed to represent the different locations across the study area.

Traditionally, Beijing has been considered as a monocentric city, and Tiananmen was often considered as the city's centre. Over recent years, the city has experienced substantial expansion. As a result, a number of sub-centres have emerged (Huang et al., 2021). Zhang et al. (2022) employed the Baidu heatmap to identify the urban centre and sub-centres in Beijing. They found four urban centres (depicted by the dark red dots in Figure 1): One is situated adjacent to Zhongguancun, which is in the northeast part of the city. The other three aligned in a horizontal direction, sequentially from left to right near Jinrongjie, Dongdan, and the CBD. Another 12 sub-centres were also found, which scattered around the inner city (captured by the orange dots).

Figure 1 Study context

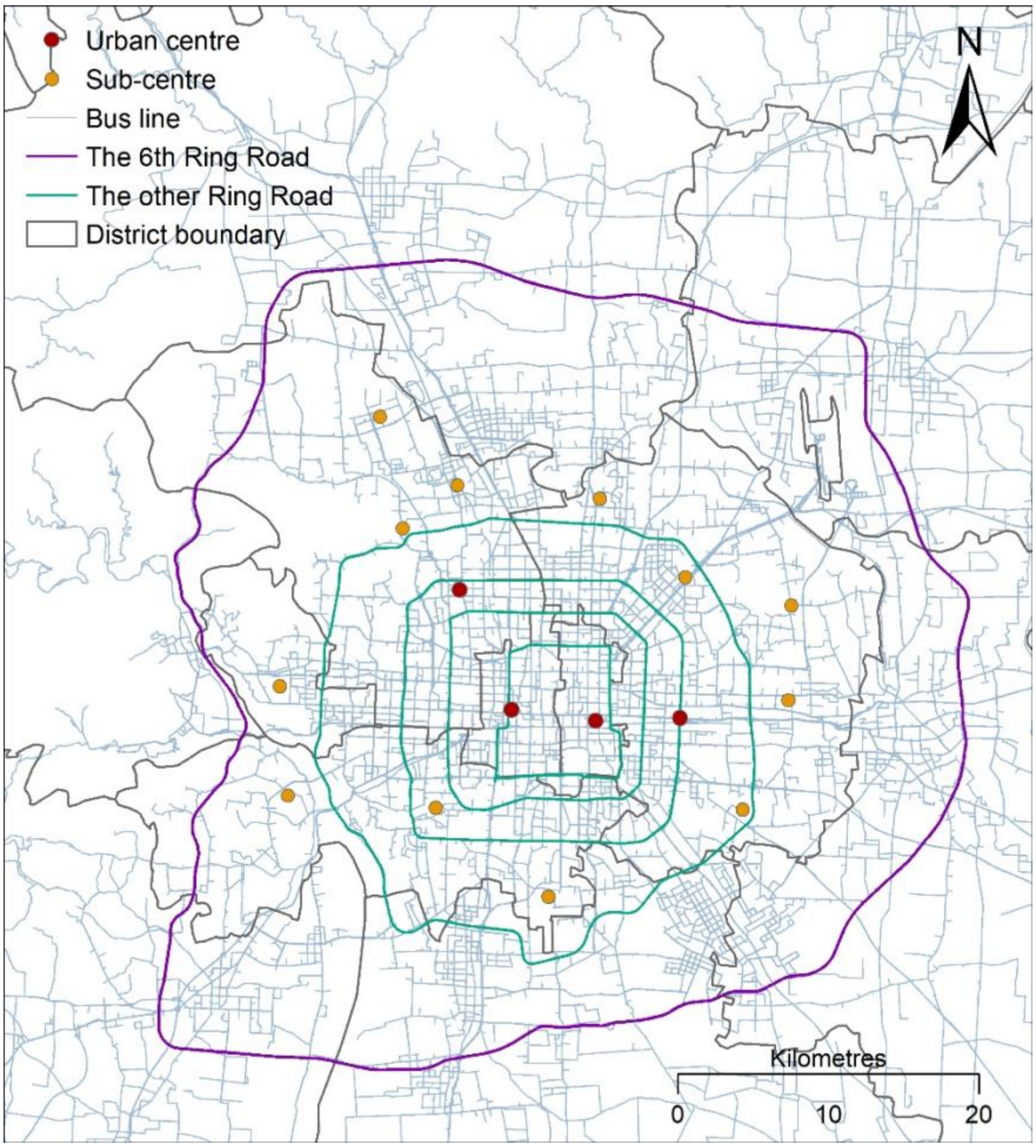

3.2 Data source

One month of SCD in June, 2016 was drawn upon to extract bus trips at the individual level. The key information includes smart card ID, date, boarding and alighting time and stops, and bus route. However, as with other types of big data, SCD lacked the personal information of the card users (e.g., gender, age, income level). Using 30 minutes as a common threshold (Jang, 2010), transfer between trip legs was identified; and trip legs were combined into bus trips. To avoid potential errors in the data, we removed the records with trip duration smaller than 1 minute and over 3 hours, which were rare in the study context. Drawing on the extracted trips, spatial and temporal variability of bus trips was estimated and investigated.

The current study also pays particular attention to trip records beginning in the morning (7-9 AM) and evening (17-19 PM) peak hours. This is due to that the demand for transit service (including bus) is most pronounced during these periods, which also posits major challenges for the operation and planning of transit systems. Trip-making during peak hours also tend to be particularly more stressful, given its more mandatory nature compared to other time periods (e.g., commuting to workplace or schools) (Zhu et al., 2019). In addition, trips within the peak hours often tend to involve relatively fixed activity anchors (e.g., home and workplace), hence allowing us to link spatial and temporal variability with given locations. The final sample size included 2,552,541 individuals with 14,217,554 trips.

Apart from SCD, other sources of data were also employed. In particular, a population data was purchased from the related department to capture the population density and demographic composition (mainly gender and age) across the city. In addition, open-source data concerning the housing estates and the built environment were extracted from a series of online open-source platforms (e.g., Baidu API, Lianjia.com—the major real estate trading agency in China). The information included the location of the main housing estates and the associated housing price, POI data for various facilities (e.g., retails, restaurants, daily services, health services and urban parks). The above datasets were used to capture the socio-demographic characteristics and built environment features. All the data were compiled at the 500-metre grid level through spatial join and interpolation. The key attributes, as such, were also estimated at the grid level.

### 3.3 Measurement of spatial and temporal variability

In line with Wei (2022), we defined the spatial and temporal variability as the spatial and temporal distance between trips made by the same person. As such, it straightforwardly tells how far or close different trips are between each other, therefore reflecting the similarity of trip-making behaviour of a given person. Based on the above definition, the variability indices were measured based on the spatial and temporal distances between trips as the underpinning components, which is illustrated in Figure 2). Specifically, for two bus trips (Trips *i* and *j*) of the same person, their main spatial and temporal features include the geographical coordinates of the origins and destinations (i.e., $x_i$, $y_i$, $x_i'$, $y_i'$ for Trip *i*, $x_j$, $y_j$, $x_j'$, $y_j'$ for Trip *j*), and the corresponding start and end of the trip (i.e., $t_i$ and $t_i'$ for Trip *i*, $t_j$ and $t_j'$ for Trip *j*).

Figure 2 Illustration of the spatial and temporal distances between bus trips

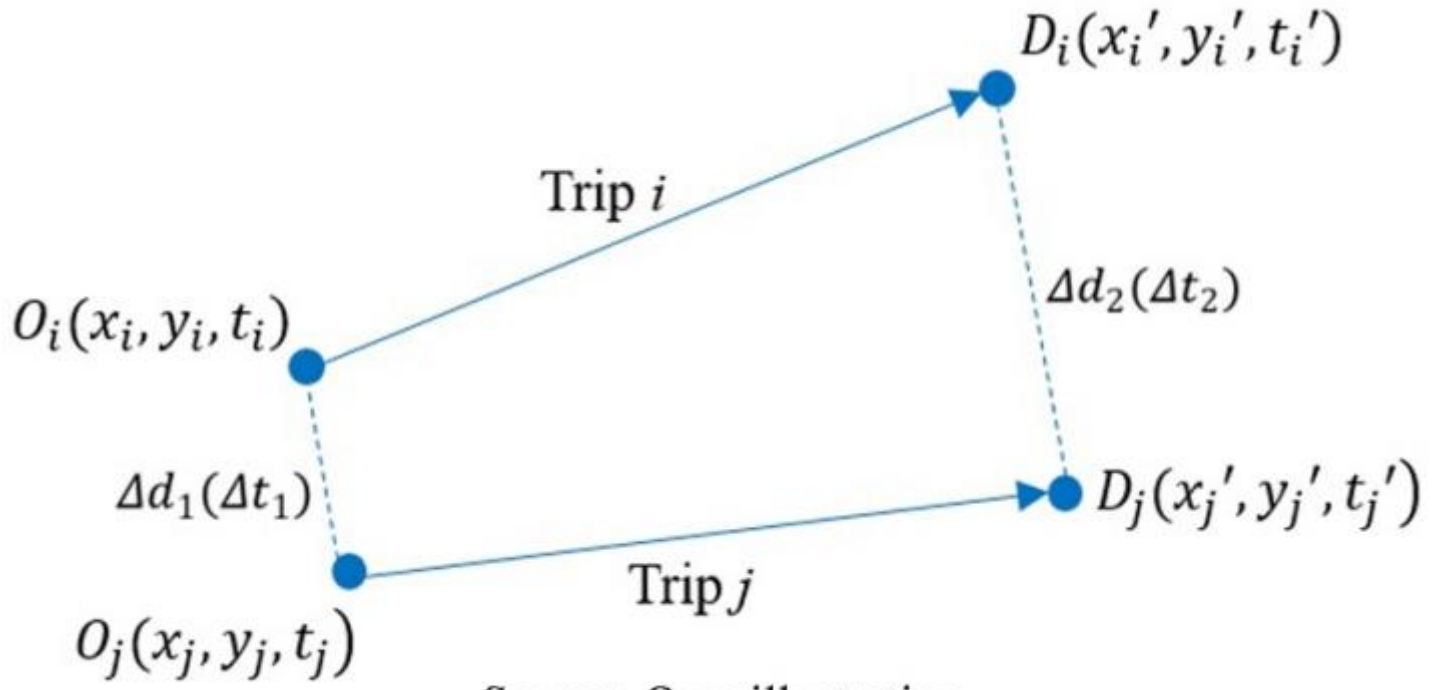

Source: Own illustration

As such, the spatial distance ($D_{ij}$) between the Trips $i$ and $j$ can be defined as the distance between the distance vectors of trip origins and destinations (i.e., $\Delta d_1$ and $\Delta d_2$) as follows:

$$D_{ij} = \sqrt{{\Delta d_1}^2 + {\Delta d_2}^2} = \sqrt{(x_i - x_j)^2 + (y_i - y_j)^2 + ({x_i}' - {x_j}')^2 + ({y_i}' - {y_j}')^2} \quad (1)$$

Similarly, the temporal distance ($T_{12}$) between Trips 1 and 2, can be defined as follows:

$$T_{ij} = \sqrt{{\Delta t_1}^2 + {\Delta t_2}^2} = \sqrt{(t_i - t_j)^2 + ({t_i}' - {t_j}')^2} \quad (2)$$

Based on the spatial and temporal distances defined above, for a given person, the spatial and temporal variability ($SV$ and $TV$) of bus trips within a certain time period can be defined as the average of the spatial and temporal distances between all pairs of trips made by the individual, respectively. The formulae for calculation can be written as:

$$SV = \frac{1}{2} * \frac{\sum_{i=1}^{n}\sum_{j=1}^{n} D_{ij}}{C_n^2} \quad (3)$$

$$TV = \frac{1}{2} * \frac{\sum_{i=1}^{n}\sum_{j=1}^{n} T_{ij}}{C_n^2} \quad (4)$$

Where $n$ is the number of bus trips, $C_n^2$ represents the number of unique trip pairs. Different from Wei (2022), we estimated the spatial and temporal variability separately. This allows us to investigate these two dimensions and their influencing factors in greater depth, hence having a more detailed understanding of the issue.

As noted above, we focused on the morning and evening peak hours (i.e., 7-9AM and 5-7PM) as the study periods. For each time period, we retained individuals with at least four trips, given that it becomes less meaningful calculating variability indices for few trips. The variability indices were estimated separately for morning and evening peak hours. A larger value of the indices indicates higher spatial or temporal variability, and vice versa.

### 3.4 Data fusion

Since the built environment and population information was compiled at the 500-metre grids, the estimated variability indices of bus travel need to be concorded with other data. To do so, we carried out the following steps.

First, for each bus rider, we identified the most frequented trip origins/destinations based on boarding and alighting stops. The extracted origins and destinations were considered as the main trip locations during the peak hours. For most cases, we expect them to be the residential location and workplace or surrounding locales. For those with more than one most frequented origins and destinations, we extracted their centroid to approximate the key activity locations. For simplicity, the extracted locations were terms as "*origins*" and "*destinations*". As such, the spatial and temporal variability of bus travel can be probed from the perspectives of both major trip origin and destination.

Second, based on the major origins and destinations, the associated variability indices of a given person were spatially joined and aggregated at the 500-metre grids. As noted above, the joined data were separated into two groups, origin- and destination-based.

Last, the mean values of the variability indices were calculated to represent the aggregate spatial or temporal variability at the neighbourhood (grid) level. A similar approach of data fusion has been applied in previous research (e.g., Zhang et al., 2021). Based on the compiled data, analysis was carried out for the major origins and destinations respectively. In this process, grids with too few observations (i.e., less than 10 individuals) were removed to ensure statistical significance. Based on the above procedures, grids with higher values mean a higher level of average spatial or temporal variability of bus use (less fixed travel locations and time), and vice versa.

### 3.5 Variable selection and model configuration

In understanding people's transit usage, the built environment and socio-demographic characteristics have been constantly identified as key influencing factors. For the former, substantial research indicated that both regional (e.g., location relative to urban centres) and local characteristics (particularly the D variables proposed by Ewing and Cervero, 2010) significantly influenced people's modal choice and transit use (e.g., ridership) (Chakour and Eluru, 2016; Diab et al., 2020). This has been attributed to that within context characterised by different built environments, the cost and convenience of using transit relative to other modes also vary significantly, hence influencing how and when people use different transport modes (Ewing and Cervero, 2010). Based on previous research, the built environment variables encompassed regional locations (distance from urban centres and subcentres), availability of transit service (stops and routes) and a series of land use attributes (e.g., density, land use mix, and road network design).

On the other hand, people's tendency and the usage patterns of transit service (e.g., trip frequency and duration) are conditioned on their socio-demographic and economic attributes, such as age, gender and income level (Tao et al., 2023; Zhang et al., 2021). These factors pertain to people's arrangement of daily activities and their access to different transport options. To capture these dimensions, we derived the proportions of female and different age groups (particularly the dependent children and older people) and the average housing price for different grids.

Last, previous research has also highlighted that the spatial and temporal variability of transit use is related to behavioural attributes, particularly including trip frequency and duration (Kim et al., 2017). Hence, we also include two behaviour features, namely the average trip frequency and average trip duration at the grid level (see Appendix I for a full list of variables). We seek to explain the spatial and temporal variability of bus use calibrating the following model:

$$VI_k = f(\beta_{tb}TB_k, \beta_{be}BE_k, \beta_{so}SO_k) \quad (5)$$

where: $VI_k$ is the variability index at location $k$, representing either spatial or temporal variability of bus use; $TB_k$ entails the two behavioural features (i.e. trip frequency and average trip duration); $BE_k$ represent a series of built environment attributes; and, $SO_k$ captures the socio-demographic and economic factors.

Model analysis was conducted for spatial and temporal variability separately, which was further distinguished based on time periods (morning versus evening peak hours) and locations (major trip origins and destinations). By doing so, we expect the results can better articulate the varying effects of relevant features. A total of eight models were estimated (four models for spatial variability and the same for temporal variability).

### 3.6 Machine learning analysis

We use a XGBoost machine learning algorithm to fit Equation (5) and identify the non-linear interactions of spatial and temporal variability of bus use as a function of behavioural, built environment and socio-demographic and economic features. XGBoost is an ensemble that combines outputs from multiple models to produce a single prediction and represents an adaptation of the gradient boosting machine algorithm proposed by Friedman (2001). As a form of gradient boosting, XGBoost utilises gradient descent to improve model performance, and decision trees are built iteratively, with each tree built to minimise the error residuals of its predecessor. XGBoost has been optimised for scalability and computational efficiency, allowing it to achieve high predictive accuracy with minimal training time (Chen et al., 2016). It has become widely recognised as one of the most effective machine learning models available and is a prominent off-the-shelf data mining technique in machine learning competitions (e.g. Chen et al., 2016).

Compared to traditional regression approaches, XGBoost has key advantages to model bus usage. XGBoost require little data engineering. It can accommodate missing data in independent variables more flexibly, while listwise deletion is used to handle missing data in regression models. XGBoost provides an appropriate balance between interpretability and model complexity in a non-parametric framework (James et al., 2013). The flexible non-parametric structure of XGBoost offers the potential to identify functional spatial and temporal structures, and novel approaches have been developed to visualise and interpret their outputs (Molnar, 2021). Furthermore, XGBoost models have successfully been used to predict overly dispersed data count in a variety of fields (e.g. Yan et al., 2010; Weng et al., 2018); that is, data with similar attributes to those often used in urban transport research. XGBoost is widely recognised to offer better prediction precision than traditional regression modelling techniques, including regression models (Ding et al., 2016) and autoregressive integrated moving average (ARIMA) models (Zhang and Haghani, 2015), as well as random forest (RF) models, support vector machine (SVM) and more complex neural networks (NN) methods (Ma et al., 2017).

Yet, we recognise that XGBoost also has limitations. XGBoost models cannot explicitly model causal relationships. They do not provide measures of uncertainty and do not test whether the difference between the relative contributions of independent variables is statistically significant. However, XGBoost models comprise a valuable framework to understand the shape and non-linearities of these associations (Rowe et al. 2022).

To analyse our XGBoost results, we use explainable machine learning techniques, specifically SHAP values. Intuitively a SHAP value quantifies the marginal contribution of each feature for a given observation to a prediction outcome. This explains how much each model feature contributes to move a model's prediction towards its final predicted value. As such, SHAP is known to represent a local view for a single data point. Unlike the commonly used Partial Dependence Plots (PDP), SHAP can handle feature interactions and complex non-linear relationships, and is theoretically bias-free, offering a more detailed and nuanced explanation of individual predictions (Lundberg et al., 2018).

## 4. Results

### 4.1 Exploratory results

An examination (see Appendix I) shows that at the individual level, a higher proportion (60-80%) of individuals had relatively lower levels of spatial (e.g., <5000) and temporal (e.g., <0.7) variability, reflecting the predominant nature mandatory trips during morning and evening peak hours (see Appendix II for the distribution of spatial and temporal variability).

Figure 3 displays the spatial patterns of the average spatial and temporal variability of bus travel at the morning and evening periods. While the spatial patterns appeared to be somewhat random at a first glance, a scrutiny indicates some distinctive patterns. In particular, for both origins and destinations, higher spatial variability was more concentrated outside the inner city areas (e.g., between the 5$^{th}$ and 6$^{th}$ Ring Roads), which was more pronounced in the southwest of the study context. On the other hand, higher temporal variability was relatively more clustered in the central areas, including areas within the 3$^{rd}$ Ring Road and certain parts in the west. Despite some localised variations of trip origins and destinations, the observed patterns display remarkably consistency across the two periods. The underlying reasons of these patterns were further explored in the modelling analysis. We also computed global Moran's Is for our variability indices (Table 1). The resulting scores were significantly positive (p value <0.01), indicating the presence of some positive clusters. Particularly, grids of high variability tend to cluster (see Figure 3).

Through correlation analysis of the average variability scores between different time periods and trip locations (not shown here), we found that spatial variability was highly correlated (Pearson correlation 0.7-0.8) with each other across time periods (morning- and evening-peak) and locations (origins and destinations), while temporal variability was moderately correlated with each other (Pearson correlation 0.2-0.4). The correlation between spatial and temporal variability was mostly weak (Pearson correlation <0.2). These results suggest that the distribution pattern was relatively consistent for spatial variability, and to a less extent, for temporal variability, while the two corresponded with each other moderately. These results indicate that the patterns of spatial variability were more consistent over space and time, but less so for temporal variability.

Table 1 Global Moran's I for the spatial and temporal variability[1]

| | Morning-origin | Morning-destination | Evening-origin | Evening-destination |
|---|---|---|---|---|
| Spatial variability | 0.239 | 0.243 | 0.240 | 0.243 |
| Temporal variability | 0.100 | 0.094 | 0.109 | 0.109 |

[1] Inverse distance was employed in estimating global Moran's I.

Figure 3 The average spatial and temporal variability at major trip origins and destinations of morning and evening periods

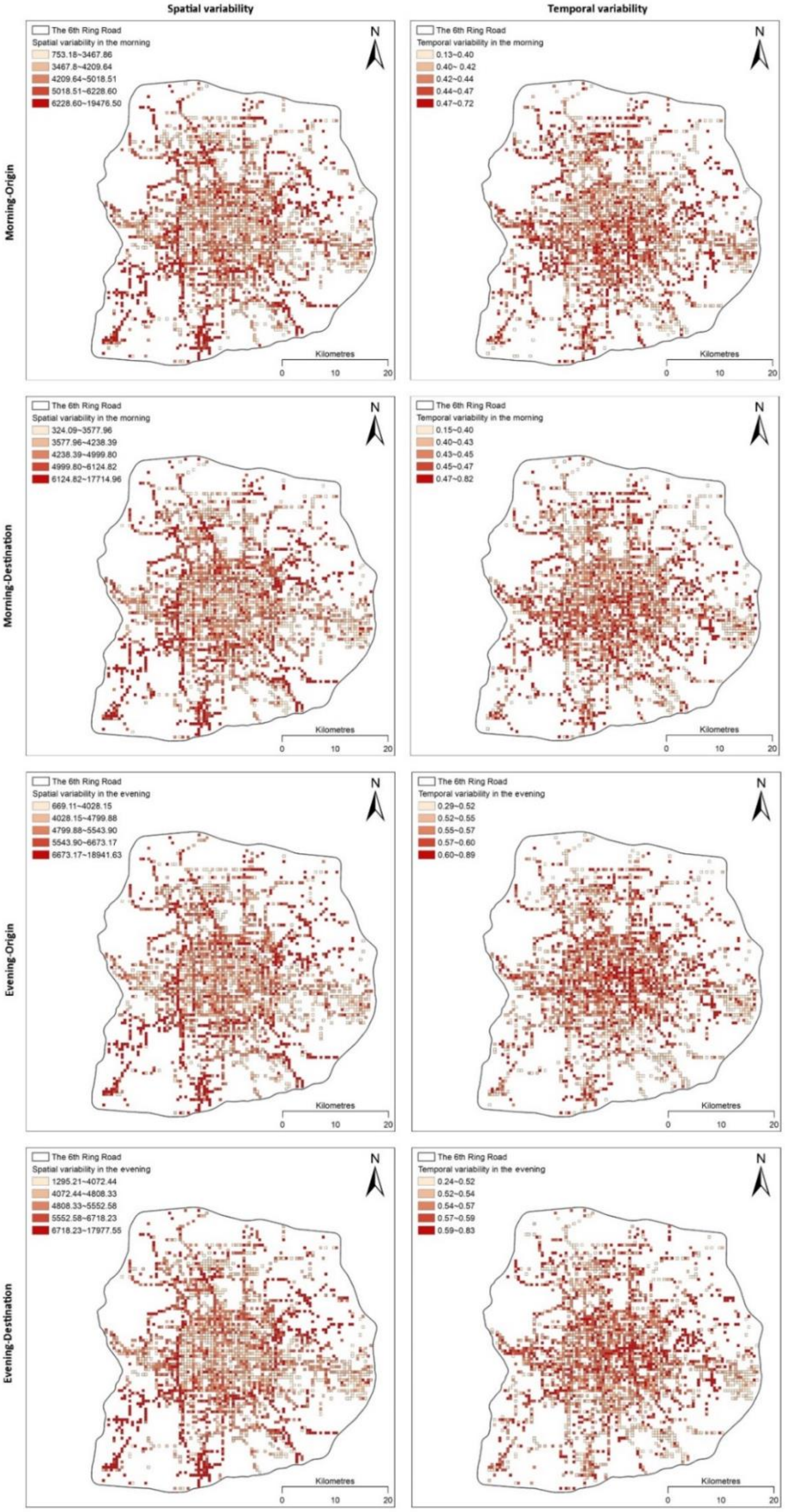

## 4.2 Relative importance of the influencing features

We estimated the relative importance (RI) of the model features based on the mean absolute SHAP values from our XGBoost model (see Table 2). The sum of all features' RIs equals 100%. Higher values of RIs indicate higher contribution of a feature in predicting spatial or temporal variability of bus use. To assist interpretation, features ranked within the top 10 were highlighted. We focused our analysis on these features.

First, the two behavioural features (i.e., average trip frequency and duration) in our models constantly displayed RIs in the top 10. Yet, these behavioural features appeared to be more important in predicting temporal than spatial variability. Furthermore, for temporal variability, the RIs for average trip duration were consistently higher than those for average trip frequency (e.g., >15%). This might relate to the link between trip duration to bus travel cost and uncertainty during busy peak hours; that is, longer trips may relate to higher travel cost and a higher chance of traffic congestion and accidents, influencing people's trip-making decisions (e.g., when to take the bus).

In terms of the built environment features, two regional location factors, namely distance to the nearest urban centre and subcentre emerged as relatively important predictors of spatial and temporal variability of bus use in Beijing. This is except for a lower RI of distance to the nearest urban centre for the temporal variability in the evening. Consistently high RIs (e.g., >10%) were found for distance to the nearest urban centre, especially for the spatial variability of bus use across the morning and evening periods. This implies that this dimension could be markedly associated with the flexibility of people's trip locations, possibly due to the more mandatory nature of trip-making during peak hours. In that case, some passengers might have to adjust their trip locations to avoid potential pitfalls (e.g., missing a bus, being late for work/school). On the other hand, the availability of both metro stations and bus stops was mostly less important, except for the temporal variability in the evening. However, the availability of bus routes was ranked within the top 10 features for all models, indicating the important role of the connectedness of bus service in shaping the variability of bus travel.

Population density exhibited higher RIs (>5%) especially in terms of spatial variability (except for morning-origin). Meanwhile, road density appeared to be relatively more important for temporal variability (except for evening-destination). The former might be due to that denser areas were also associated with more bus services. The latter finding suggests that road network configuration could influence the flexibility of bus trips more. Concerning POI-based features, POI entropy emerged as more important in relation to temporal variability, but less for spatial variability. This implies that mixed land use might induce different types of activities at trip origins and destinations, hence contributing to changed patterns of bus use. By comparison, the availability of specific facilities all exhibited low RIs (e.g., mostly between 1-3%).

With regard to socio-economic and demographic features, housing price emerged as more important for all models. This reflects that difference in socio-economic resources could also influence people's choice of flexible travel. The proportion of females was also more linked to the spatial variability of bus travel than for the temporal variability, indicating the presence of gender difference in daily mobility. The proportions of independent children and older people also emerged as relevant in most models. This can be attributed to that the presence of these groups might either lead to distinct travel patterns at origins, or attract supply of relevant services and facilities at destinations, thereby inducing more varied travel patterns by bus.

Table 2 Relative importance of the influencing features (%)

| | Abbreviation | Spatial variability | | | | Temporal Variability | | | |
|---|---|---|---|---|---|---|---|---|---|
| | | Morning-origin (N=2536) | Morning-destination (N=2608) | Evening-origin (N=2658) | Evening-destination (N=2665) | Morning-origin (N=2536) | Morning-destination (N=2608) | Evening-origin (N=2658) | Evening-destination (N=2665) |
| Behavioural features | | | | | | | | | |
| Average trip frequency | tripfreq | **5.52** | **6.28** | **4.89** | **6.69** | **12.79** | **8.21** | **13.15** | **20.89** |
| Average trip duration (hour) | avedur | **17.66** | **11.08** | **6.66** | **13.19** | **19.06** | **18.65** | **28.57** | **24.49** |
| Built environment features | | | | | | | | | |
| Distance to the nearest urban centre (km) | centdist | **16.43** | **16.85** | **18.27** | **16.27** | **5.36** | **6.70** | 3.08 | 2.30 |
| Distance to the nearest subcentre (km) | subcendist | **4.41** | **7.24** | **6.51** | **5.40** | **5.34** | **5.64** | **6.10** | **3.97** |
| Availability of metro stations | metrosta | 2.61 | 0.73 | 1.05 | 0.91 | 4.01 | 2.90 | **3.79** | 0.85 |
| Availability of bus stops | busstop | 1.78 | 2.12 | 2.67 | 2.22 | 3.17 | 1.24 | 0.95 | **4.44** |
| Availability of bus routes | busroute | **8.87** | **7.47** | **8.32** | **8.04** | **5.37** | 4.54 | **4.28** | **5.01** |
| Population density (persons/km$^2$) | popden | 3.77 | **5.35** | **4.92** | **5.45** | 3.40 | **4.87** | 2.13 | **3.50** |
| Road density (km/km$^2$) | roadden | **4.60** | 3.93 | 4.04 | 3.41 | **4.33** | **7.16** | **4.38** | 2.96 |
| POI entropy | poi_entro | **4.20** | 3.88 | 3.06 | 3.62 | **4.38** | **4.95** | 3.11 | **3.43** |
| Availability of restaurants | eat | 1.81 | 1.70 | 3.91 | 1.60 | 2.00 | 3.37 | 1.42 | 2.67 |
| Availability of recreational facilities | recrea | 2.33 | 2.29 | 2.55 | 2.92 | 1.86 | 3.14 | 0.94 | 1.53 |
| Availability of daily services | dailyser | 1.79 | 2.26 | 2.43 | 2.97 | 2.53 | 4.17 | 1.54 | 2.06 |
| Availability of financial facilities | finan | 2.05 | 1.78 | 0.73 | 1.82 | 2.97 | 2.52 | 0.83 | 2.48 |
| Availability of cultural facilities | cult | 1.77 | 2.79 | 2.42 | 2.51 | 2.15 | 2.07 | 1.84 | 0.73 |
| Socio-demographic and economic features | | | | | | | | | |
| Housing price (RMB) | houseprice | **4.85** | **5.41** | **5.77** | **7.44** | **4.32** | **5.00** | **7.04** | **3.96** |
| Proportion of female (%) | female | **6.81** | **9.01** | **10.12** | **6.80** | 3.46 | 4.37 | **3.63** | **5.67** |
| Proportion of dependent children (%) | juveni | **4.70** | **4.28** | **5.50** | **4.55** | **5.78** | **5.87** | **4.38** | 2.59 |
| Proportion of older people (%) | oldage | 4.03 | **5.55** | **6.17** | **4.19** | **7.71** | **4.64** | **8.85** | **6.47** |

2

4.3 Non-linear relationships for behavioural features

Next, we examined the non-linear relationships using the SHAP dependence plot. The dependency plot illustrates the distribution of the SHAP values (on the vertical axis) against the feature values (on the horizontal axis). By doing so, it can reveal the changes in the influence of a given feature. In the plots, the sign of the SHAP values indicate the direction of influence (i.e., positive or negative), while larger absolute values indicate stronger effects, and vice versa. For this part of analysis, we focused more on the behavioural features and built environment characteristics. This is due to that these dimensions often constitute the main targets of related policymaking and interventions.

Concerning the spatial dimension (Figure 4), a glimpse shows that those with lower trip frequency and shorter trip duration on average also demonstrated lower spatial variability, and vice versa. Yet, there exist some distinctions. Specifically, when average trip frequency was below 8, its effect on spatial variability were mainly negative. As the trip frequency was between 8 and 10, its effect turned positive and the spatial variability was also higher. For average trip duration, a linear pattern was observed for morning destination, whilst some turning points were present for the other three models. For the former, lower average trip duration was linked to lower spatial variability (i.e., negative effects), and vice versa. For the other three models, turning points can be observed around 0.4-0.5 hour, beyond which the positive effect of the trip duration remained relatively constant or weakened (i.e., for morning origin).

Figure 4 SHAP dependence plots for the behavioural features for the spatial variability of bus use

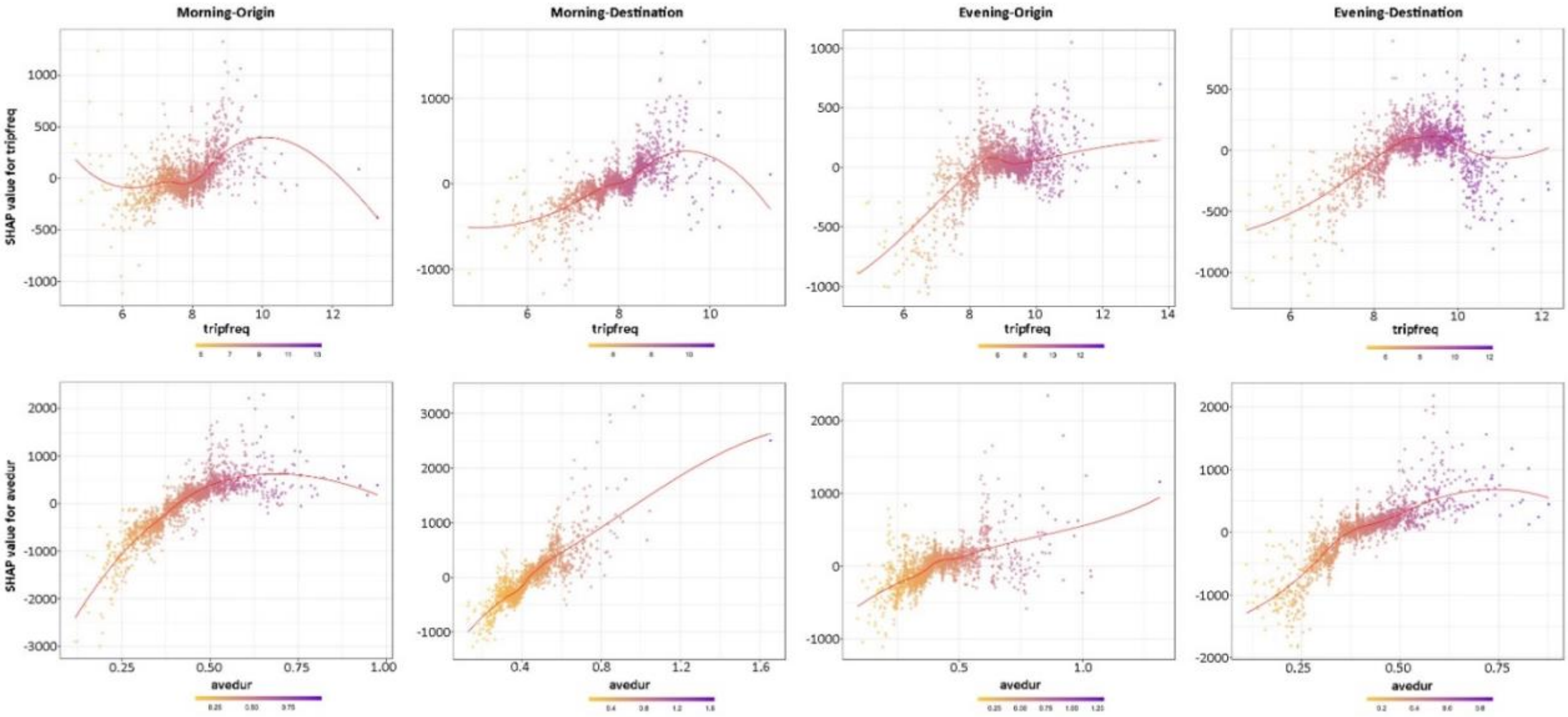

Turning to temporal variability (Figure 5), the revealed relationships appeared to be relatively similar for both morning and evening periods, and origins and destinations as well. For average trip frequency, a convex pattern can be identified for all models. When the trip frequency was both lower (<7) and higher (>8), its effect on the temporal variability was mainly negative. Yet, for considerably higher trip frequency (>10), the negative effect was also more pronounced compared to the lower frequency. This finding might mainly reflect the travel patterns of commuters who had relatively fixed schedules. On the other hand, the revealed effects of average trip duration mainly demonstrated a more straightforward distribution. As the trip duration was below 0.5 hour (or 0.4 hour for morning- and evening-destination), its effect on temporal variability was mainly negative, while turning positive for longer trip duration. This implies that compared to shorter trips, longer trip duration on average was associated with stronger fluctuation of trip time.

Figure 5 SHAP dependence plots for the behavioural features for the temporal variability of bus use

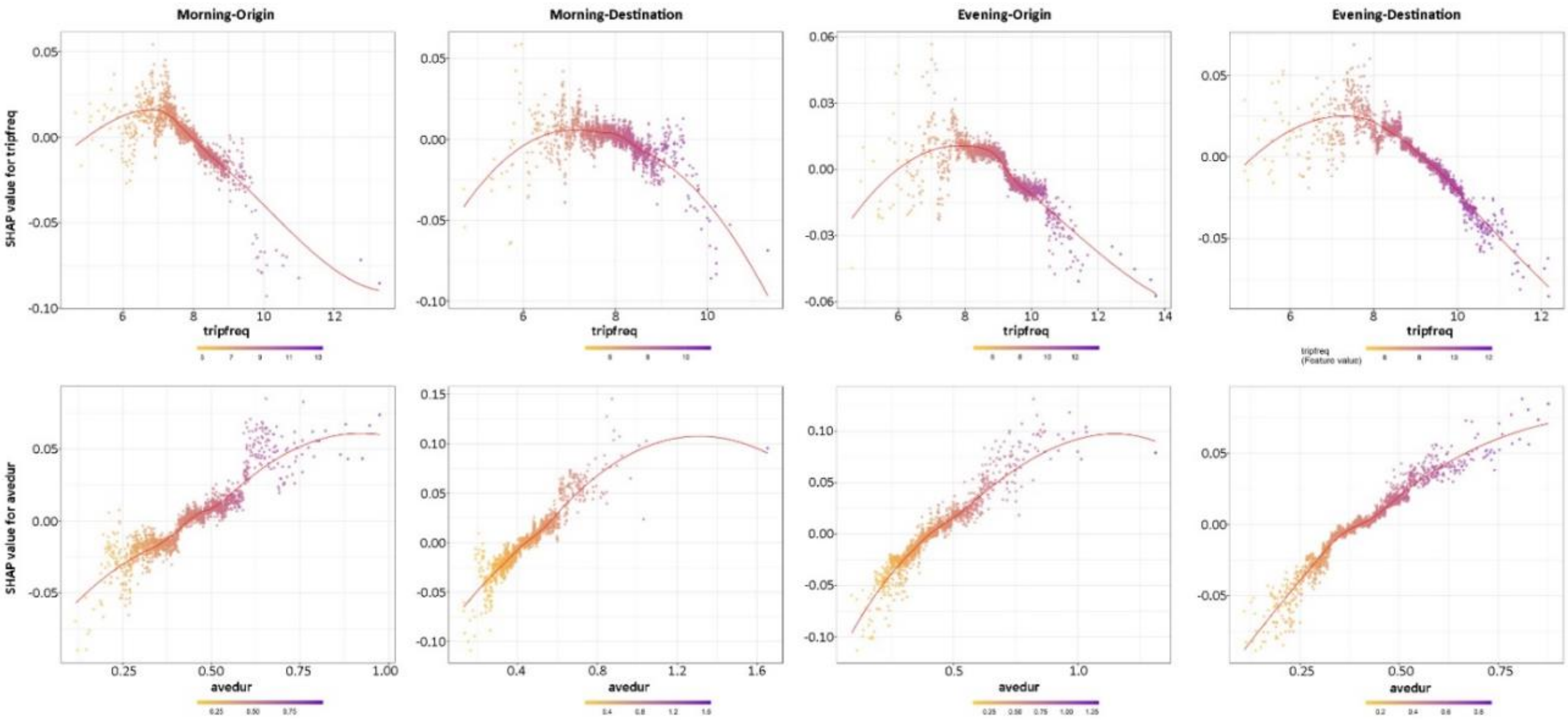

## 4.4 Non-linear relationships for built environment features

We found that regional factors, availability of bus routes, population density, road density and POI entropy were constantly among the top 5 most important built environment features for most models. Hence, for this part of analysis, we focused on these features to enable a more focused discussion. For enhanced clarity, plots for the same variable were presented in the same row. In cases where a given feature was not within the top 5, a note reading “NOT in top 5” was placed instead. Based on these rules, Figure 6 shows the dependency plots for spatial variability of bus trips.

First, the effects of distance to the nearest urban centre and subcentre exhibited marginal or moderately negative effects at lower values (e.g., <7-8 km). At higher values (e.g., >10 km), their effects appeared to be divergent (see Figure 6). Specifically, the effect of distance to the urban centre turned increasingly positive as it grew larger (or increasing spatial variability), whilst markedly negative effects (or lowering spatial variability) were observed for larger distance to the subcentre. Such opposite effects imply that proximity to urban centres and subcentres acted differently in shaping people's trip locations by bus. Particularly, a larger distance from the urban centres might force people to take different bus services (e.g., using different routes or stops) to fulfil their peak hour trips, possibly because of lower density and more dispersed land use that were less convenient for bus use. On the other hand, as subcentres are located mainly in the suburban areas, a larger distance from them might compel people to shift to other modes (e.g., private cars, metro), while the remaining captive users mainly used bus for mandatory trips with relatively fixed locations (e.g., between home and workplace).

The effects of the availability of bus routes were largely consistent across time periods and locations. As it was at a lower value (e.g., around 20), its effect was mainly negative, hence reducing spatial variability of bus trips. Yet, as it became larger (e.g., >30), its effect turned positive, although the magnitude of the effect increased subtly at higher values. This suggested while more bus stops available might contribute to higher flexibility of bus usage, there existed a potential threshold beyond which the return diminishes.

Population density ranked the top 5 for morning-destination and the evening period, but not for morning-origin. For the three models, it appeared the effect of this feature was positive (or increasing spatial variability) mainly at a lower value (e.g., around 10000 persons/km$^2$), but largely marginal afterwards. This implies that areas with lower density might to some extent also force people to use different bus services. By comparison, road density at both lower (around 10 km/km$^2$) and higher values (>25 km/km$^2$) was associated with a positive effect on spatial variability, showing a concave pattern. This is especially the case for the morning period (and for both origins and destinations). This, again, might involve different mechanisms: lower road density could lead to a forced level of mobility in using bus service (e.g., walking an extra distance to different bus stops), whilst higher road density might increase people's flexibility in accessing bus service. For evening-origin, only lower road density was associated with heightened spatial variability.

Last, POI entropy ranked among the top 5 for morning-origin and evening-destination. Considering the tidal patterns of peak hour travel, the origins and destinations of the two time periods might to some extent coincide. Discernibly, negative effects of this feature were mainly observed for higher values (e.g., over 2), indicating that more mixed land use was linked to lower spatial variability of bus trip. A plausible explanation is that in areas with varied facilities in vicinity, people might prefer active travel (e.g., walking) to taking buses due to enhanced accessibility.

Figure 6 SHAP dependence plots for the built environment features for the spatial variability of bus use

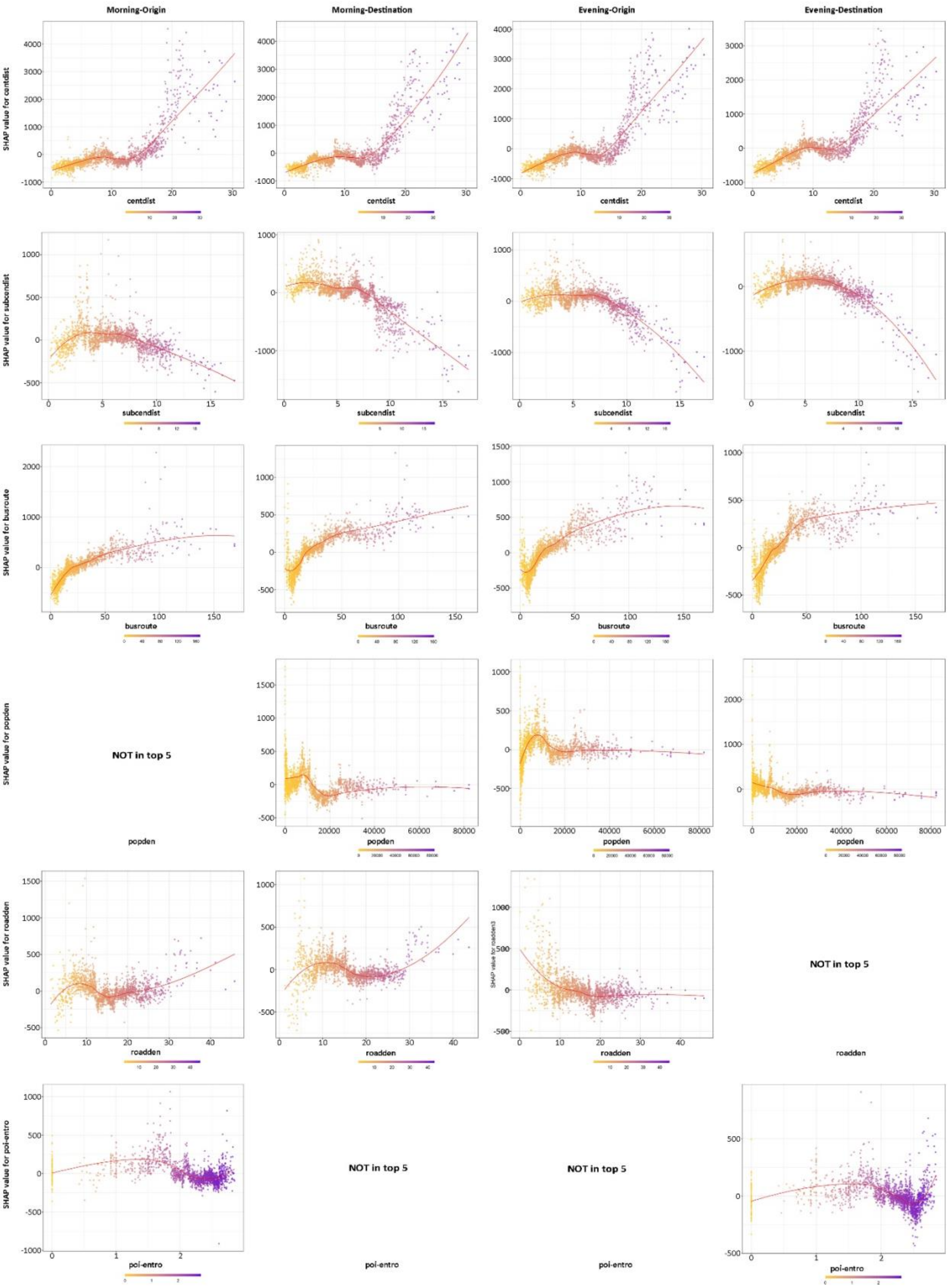

Figure 7 displays the dependency plots for temporal variability in relation to the built environment features. Regarding regional features, distance to the nearest urban centre entered the top 5 for three models except for evening-destination. For the morning period, lower values (<10 km) of this feature were mainly linked to positive effects (or increasing temporal variability of bus use), whilst negative effects (or reducing temporal variability) were more observed for higher values (>10 km). This suggests the being away from the urban centres to some extent reduced the flexibility of trip time by bus. However, for the origins in the evening period, higher values of this feature (>16 km) captured more positive effects. This might reflect the increasing uncertainty of bus travel in the evening.

Distance to the nearest subcentre also ranked among the top 5 in all four models. For the morning period, the effects of this feature changed from slight positive to moderately negative as its value increased, which appeared to be consistent for both origins and destinations. This hints that larger distance to subcentres might to some extent reduce the flexibility of bus travel as well. By comparison, in the evening period, larger values of this feature (>10 km) corresponded to negative effects (or decreasing temporal variability) that were strong than in the morning. Again, similar to the above findings of spatial variability, being away from the subcentres might induce more shift from bus to other modes (e.g., metro, private cars) in the evening, while the remaining riders were probably captive users who had relatively inflexible travel time.

In three models (except for morning-destination), the availability of bus routes emerged as more important. For this feature, positive effects mainly concentrated within a lower range (around 20), beyond which marginal or weak negative effects were observed. This suggests that fewer bus routes available might also add uncertainty to bus travel in peak hours. Concerning density variables, marked fluctuations of effects were observed mainly for lower ranges of population density (<10000 persons/km$^2$) in morning- and evening-destinations, indicating possibly increased uncertainty of bus services in low-density areas. On the other hand, for road density, notable positive effects mainly existed in lower values of this feature (<10 km/km$^2$). A possible explanation is that compared to denser road network, a sparser network layout might make accessing bus stops less convenient, hence adding uncertainty to travel time.

Last, the effects of POI entropy on temporal variability appeared to vary between time periods and locations. For morning-origin, its effect was mainly positive for higher values, indicating possibly more flexible bus use patterns due to more mixed functions. Yet, for morning-destination and evening-origin, its effect first turned positive, and shifted to slightly negative at higher values. These nuanced changes indicate that mixed land use induced more flexible bus use only within a certain range. Such a turning point was even more pronounced for evening-destination. A plausible speculation is that for return trips in the evening, people might not prefer environments with highly mixed facilities, which can be associated with increasing crowdedness or noise. Rather, they might return home quickly as a routine.

Figure 7 SHAP dependence plots for the built environment features for the temporal variability of bus use

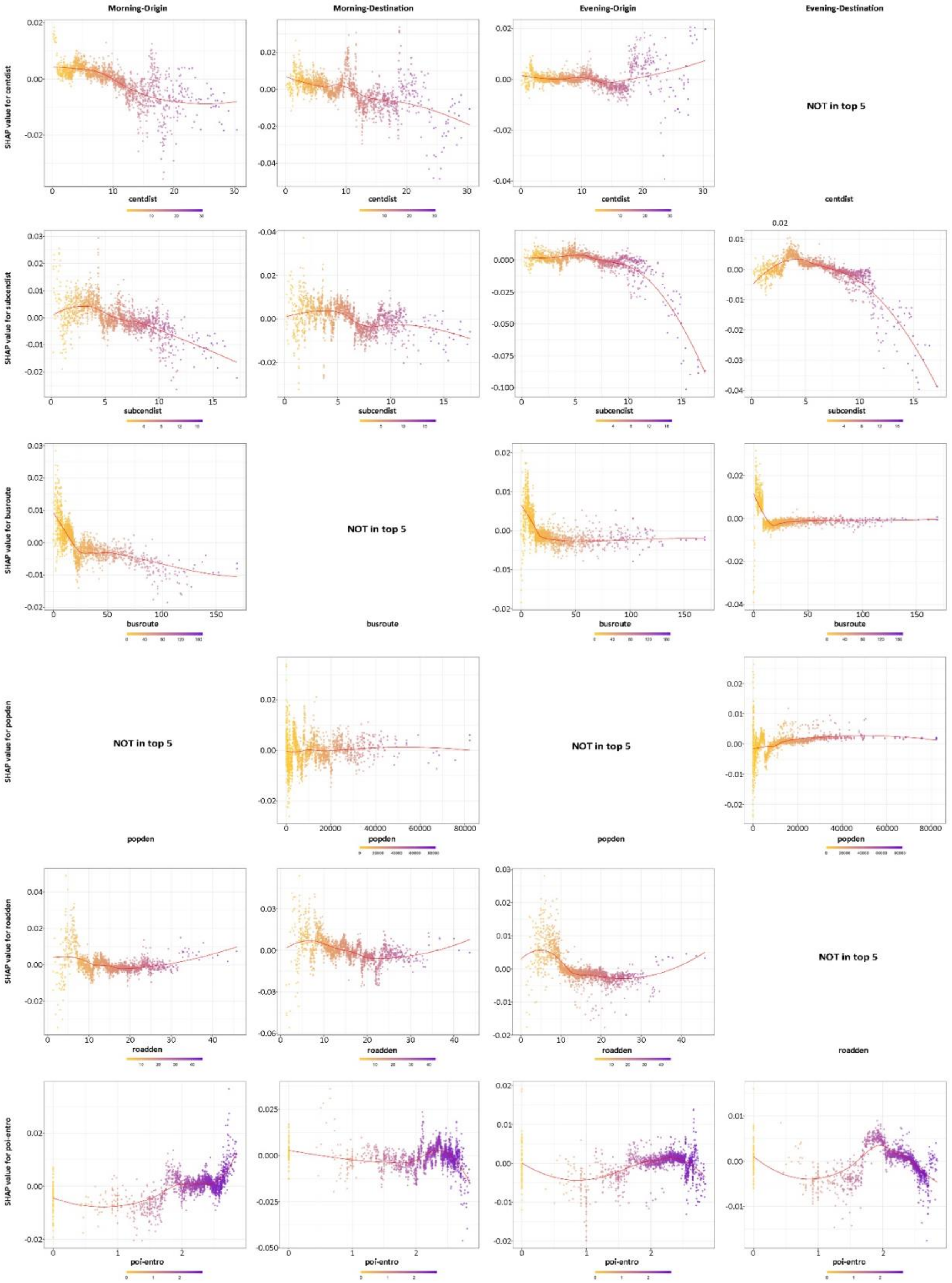

## 5. Discussion

The current study presents an in-depth investigation of the spatial and temporal variability of bus travel in Beijing. It provides some first analysis seeking to disentangle the underlying predictive features of the spatial and temporal variability of bus use. Our analysis indicated that two trip-making features—trip frequency and duration, are among the most important predictors of spatial and temporal variability of bus use. Using explainable machine-learning, we showed the shape of complex, non-linear relationships in the association between the spatial and temporal variability of bus use and built environment features. We identified key threshold points indicating sudden changes in the spatial and temporal variability of bus use in Beijing as the local availability of bus routes and density of built environment features increase. Such insights have provided importantly new perspectives in understanding and managing transit use.

First, our findings suggest that the spatial and temporal variability of bus use reflect behavioural features. Our evidence indicates that lower trip frequency and average trip duration were negatively associated with spatial and temporal variability, while reverse was noted for average trip duration (see Figures 4 and 5). We also identified inflection points with the relationship between trip frequency and the temporal variability of bus usage shifting to negative at higher values (see Figure 5). This reflects regular and more similar temporal patterns for frequent bus riders. Furthermore, the positive effects of average trip duration on the spatial variability weakened at higher values especially for morning-origin and evening-destination (see Figure 4). This implies that some bus riders might seek to minimise the uncertainty associated with longer trips or travel to specific areas when they travel over long distances.

Our analysis further unveiled the non-linear relationships between our variability indices and built environment features. For spatial variability of bus use, distance to the nearest urban centre and subcentre displayed pronounced yet divergent effects at higher ranges (see Figure 6), suggesting different behavioural responses of bus riders travelling to more remote locations within Beijing. Higher availability of bus routes showed a positive effect on the spatial variability, but only to a limited extent. Additionally, the negative effects of low population and road density on the spatial variability might capture the outcomes of less friendly environment for bus usage. Higher POI entropy, on the other hand, appeared to reduce the spatial variability of bus use, hinting possible modal shift in such circumstances. For temporal variability, larger distance to the nearest urban centre and subcentre was constantly linked to negative effects, also implying potentially less convenient conditions for bus use. Meanwhile, the effects of the availability of bus routes and density on temporal variability mainly fluctuated at their lower ranges (see Figure 7). These findings suggest that fewer bus routes and travelling between low-density areas might add uncertainty to travel time for specific trip times and locations.

Taken together, these empirical findings have implications for the development of more responsive operation and planning of transit service. First, our study affirms that transit agencies may combine the conventional and variability indices to better distinguish passenger groups with distinctive travel demands in space and time, informing more targeted marketing strategies (e.g., providing more customised transit information and services) (Egu and Bonnel, 2020; Kieu et al., 2015). Second, through probing into the non-linear relationships, we found that the spatial and temporal variability of bus use were possibly subject to varying "push" and "pull" forces of the built environment. Specifically, greater distance to urban centres and subcentres might not only prolong trips, but also add uncertainty and inconvenience to commute trips, thereby contributing to (potentially involuntary) spatial variability while reducing flexible trip-making time. On the other hand, higher availability of bus routes coupled with denser road network might enable more flexible bus use during peak hours.

Drawing on these findings, transit agencies may initially evaluate whether the variable patterns of transit use can be beneficial or not (e.g., reliving commuting stress or requiring extra effort for bus use). Based on such a knowledge, more targeted operation and management of bus services can be considered, for example, installing more devices for real-time information (e.g., routes and schedule), providing more customised information for commuters, adding more flexible mini-bus services in peripheral areas. Furthermore, the revealed threshold effects of the built environment characteristics can provide an evidence base for planning measures or interventions to allow more flexible use of time for bus trips and related activities enroute. This may entail densifying road networks and controlling the types and quality of varied facilities (e.g., grocery shopping) around bus stops.

Despite the new insights, the current research also has some limitations that form avenues for future research. First, the measurements of spatial and temporal variability were solely based on the geo-locations and timestamps of origins and destinations, as they arguably represent the most important anchors of trip-making. Yet, future research may seek to refine the measurement by also taking account of the mid-points of the trips, thereby more accurately gauging trip similarity (Shen, 2019). This can also serve the based for extended research topics, e.g., assessing the variability of activity space. Second, the current study only focused on bus as a main transit mode within the study context. Future research may expand the scope by including alternative modes such as the metro, private vehicles and shared biking. While this will provide a fuller picture of daily travel patterns, integration of different data sources may become necessary and posit a major challenge. Third, from the empirical findings we consider that variability of daily trip-making may not be entirely voluntary and contribute to more flexible time use. Future research may continue this line of research by looking into the perceptions and experience of variability in daily mobility, hence providing more concrete evidence and implications. Finally, due to data limitation, we investigate the variability of bus use at an aggregate level. Future research may seek ways to complement more personal information and further investigate how different personal features may affect the variability of transit use. It would be also of interest to explore the spatial and temporal variability of different trip purposes and non-peak periods to gain a more comprehensive picture.

## 6. Conclusions

A fuller grasp of how spatially and temporally sticky daily public transit use constitutes a critical component of understanding human mobility dynamics in cities and devising more responsive and customised transit services. We contribute to the literature by (1) developing simple yet robust metrics to quantify the spatial and temporal variability of bus use, and (2) identifying the key predictive factors contributing to the spatial and temporal variability of bus use and their non-linear relationships using interpretable machine learning. Results highlight threshold points in the spatial and temporal variability of bus usage reflecting differences in travel behavioural patterns and in the density and availability of key built environment features. In particular, the spatial and temporal variability of transit use displays a convex shape, with increased trip frequency, reflecting that both infrequent and frequent users might demonstrate lower flexibility in their bus travel. A negative association between the temporal variability of bus use, and greater distance to urban centres and subcentres indicates a certain level of forced regularity in transit use in more peripheral areas, possibly due to the added uncertainty, inconvenience and constraints of the transport network in such places (e.g., sparser, less accessible stops and reduced long-trip options). At the same time, higher availability of bus routes and a denser road network are associated with high spatial and temporal variability reflecting added flexibility to bus usage by providing more convenience and flexibility in trip-making. These findings can inform transport and operation plans for enabling a more varied offer of bus services and helping to develop a more resilient transport network. Future work could extend our analysis by identifying differences in the spatial and temporal variability of bus usage across population subgroups and investigating different city contexts. Such work would enable developing a more systematic understanding of bus use patterns and identifying regularities in the patterns of spatial and temporal variability in bus usage. By doing so, a more concrete evidence and meaningful discourse can be formed to guide future transit planning in an era of increasing motorisation.

**Appendix I Descriptive statistics for the key features**

| Mean (std. deviation) | Abbreviation | Morning-origin (N=2536) | Morning-destination (N=2608) | Evening-origin (N=2658) | Evening-destination (N=2665) |
|---|---|---|---|---|---|
| Variability indices | | | | | |
| Spatial variability | - | 5496.135 (1946.678) | 5545.737 (2008.568) | 4998.704 (1998.087) | 4949.34 (1832.145) |
| Temporal variability | - | 0.559 (0.061) | 0.555 (0.056) | 0.436 (0.054) | 0.438 (0.060) |
| Behavioural features | | | | | |
| Average trip frequency | tripfreq | 7.84 (0.77) | 7.89 (0.71) | 9.00 (0.98) | 8.93 (1.02) |
| Average trip duration (hour) | avedur | 0.44 (0.11) | 0.44 (0.12) | 0.41 (0.13) | 0.41 (0.11) |
| Built environment features | | | | | |
| Distance to the nearest urban centre (km) | centdist | 9.45 (5.74) | 9.58 (5.82) | 9.65 (5.82) | 9.58 (5.79) |
| Distance to the nearest subcentre (km) | subcendist | 5.84 (2.97) | 5.88 (3.00) | 5.88 (3.00) | 5.88 (3.00) |
| Availability of metro stations | metrosta | 0.85 (0.98) | 0.84 (0.98) | 0.84 (0.98) | 0.84 (0.98) |
| Availability of bus stops | busstop | 3.40 (2.48) | 3.69 (2.48) | 3.68 (2.48) | 3.68 (2.47) |
| Availability of bus routes | busroute | 27.06 (24.59) | 27.06 (24.73) | 26.72 (24.69) | 26.68 (24.49) |
| Population density | popden | 62986 (31051) | 62676 (31159) | 62206 (31177) | 62497 (31039) |
| Road density (km/km$^2$) | roadden | 15.48 (6.49) | 15.40 (6.48) | 15.34 (6.47) | 15.40 (6.45) |
| POI entropy | poi_entro | 2.10 (0.61) | 2.10 (0.61) | 2.09 (0.62) | 2.09 (0.62) |
| Availability of restaurants | eat | 28.43 (41.88) | 28.13 (41.51) | 27.74 (41.31) | 27.93 (41.44) |
| Availability of recreational facilities | recrea | 46.00 (132.29) | 45.05 (130.48) | 44.34 (129.27) | 44.93 (129.62) |
| Availability of daily services | dailyser | 46.13 (52.22) | 45.64 (51.77) | 44.90 (51.70) | 45.33 (51.87) |
| Availability of financial facilities | finan | 5.42 (11.54) | 5.33 (11.40) | 5.24 (11.32) | 5.29 (11.34) |
| Availability of cultural facilities | cult | 11.88 (17.76) | 11.75 (17.66) | 11.54 (17.51) | 11.62 (17.49) |
| Socio-demographic and economic features | | | | | |
| Housing price (RMB) | houseprice | 63771 (21621) | 63409 (21562) | 63236 (21527) | 63435 (21574) |
| Proportion of female (%) | female | 46.02 (4.93) | 46.00 (4.92) | 45.94 (4.97) | 45.98 (4.95) |
| Proportion of dependent children (%) | juveni | 2.72 (0.95) | 2.73 (0.98) | 2.73 (0.98) | 2.72 (0.96) |
| Proportion of older people (%) | oldage | 5.50 (2.42) | 5.46 (2.42) | 5.43 (2.42) | 5.46 (2.41) |

**Appendix II The distribution of spatial and temporal variability of bus use at the individual level**

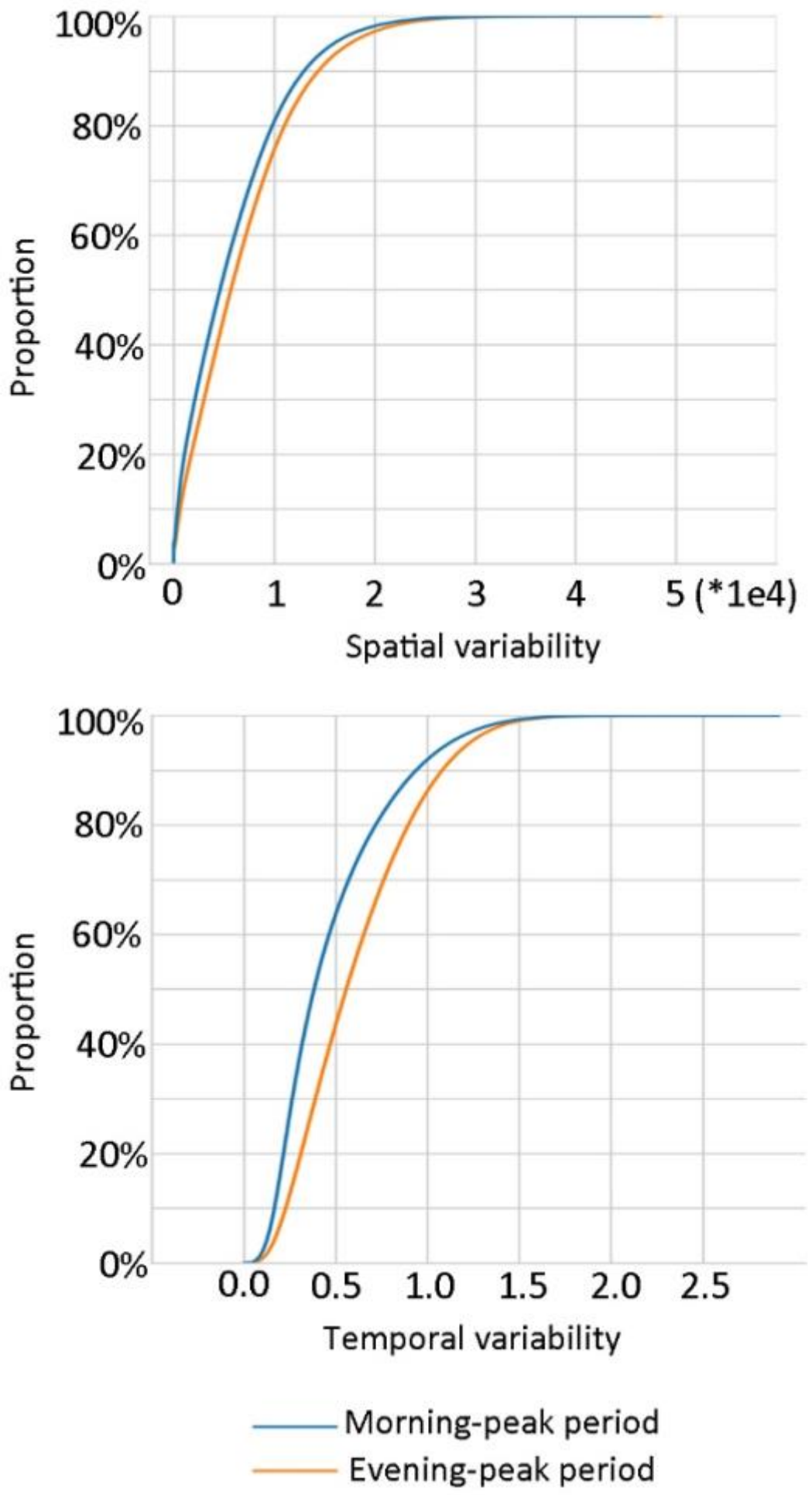